\documentstyle[prl,aps,epsf]{revtex}
\newcommand \beq{\begin{eqnarray}}
\newcommand \eeq{\end{eqnarray}}
\newcommand{\set}[2]{\newcommand{#1}{#2}}
\set{\pa}{\partial \over \partial\, }
\set{\leftvector}{\stackrel{\leftarrow}{\partial }}
\set{\rightvector}{\stackrel{\rightarrow}{\partial }}
\begin{document}
\twocolumn[\hsize\textwidth\columnwidth\hsize
           \csname @twocolumnfalse\endcsname
\title{Giant Octupole Resonance Simulation}
\author{Rainer Walke, Klaus Morawetz}
\address{Fachbereich Physik, Universit\"at Rostock, D-18051 Rostock,
Germany}
\maketitle
\begin{abstract}
Using a pseudo-particle technique we simulate
large-amplitude isoscalar giant octupole excitations
in a finite nuclear system. Dependent on the 
initial conditions we observe either clear octupole modes or
over-damped octupole modes which
decay immediately into quadrupole ones. This shows clearly a behavior beyond linear response. We propose that octupole modes
might be observed in central collisions of heavy ions.  
\end{abstract}
\vskip2pc]
%
%
Giant resonances regain much attention presently for the investigation of many particle effects in
finite quantum systems. While most of the theoretical treatments of oscillations rely on the linear response method or RPA methods, large amplitude oscillations require methods beyond.  Especially the question of the appearance of chaos is
recently investigated \cite{BGZS94,BurBal95,Mor97}. 
The hypothesis was established that the octupole mode is overdamped due to negative curved surface and consequently additional chaotic damping
\cite{JarSwi93,BloShi93,BloSka97}. Here we want to discuss at which conditions one might observe octupole modes at least in Vlasov - simulation of giant resonances.

We will consider different initial conditions of isoscalar giant 
resonances using a pseudo-particle simulation of Vlasov kinetic
equation\cite{BerDas88}. 
%
%
With a local potential $U({\bf r})$ the quasi-classical Vlasov equation reads
\begin{eqnarray}
\frac{\partial}{\partial t}f({\bf p,r,}t) + \frac{{\bf p}\hbar}{m}{\bf \nabla_r}f -
 \frac{1}{\hbar}
 {\bf \nabla_p}f \,{\bf \nabla_r}U({\bf r}) &=& 0 \ .
\label{e1}
\end{eqnarray}
We represent the distribution function $f({\bf p,r},t)$ by a sum of
pseudo-particle distributions
\begin{eqnarray}
f({\bf p,r},t) \approx f_0({\bf p,r},t) &=& \sum\limits_{i=1}^{A N} 
      \frac{1}{N} f_S({\bf p}-{\bf p_i}(t),{\bf r}-{\bf r_i}(t)) \; 
\label{2}
\end{eqnarray}
and use Gaussian pseudo-particles 
\begin{eqnarray}
f_{S}({\bf p-p_1,r-r_1}) = c\,\; e^{-({\bf p-p_1})^2/2\sigma_p^2}\;\; e^{-({\bf r-r_1})^2/2\sigma_r^2}
\label{3}
\end{eqnarray} 
at ${\bf r_1}$ with momentum ${\bf p_1}$ \cite{Wong82}.
These pseudo-particles follow 
classical Hamilton equations
\begin{eqnarray}
\hbar\dot{\bf p}_i&=&-{\bf \nabla} U,\; \; \; \dot{\bf r}_i=\frac{\hbar {\bf p}_i}{m} \; .
\end{eqnarray}
We assume for the interacting nucleons a phenomenological density
dependent Skyrme interaction\cite{ColDiT95} which results into the
mean field 
  \begin{eqnarray} 
    U(\varrho) &=& a \frac{\varrho}{\varrho_0} 
      + b \left (\frac{\varrho}{\varrho_0}\right)^s
  \end{eqnarray}
with $a=-356{\rm MeV}$, $b=303{\rm MeV}$ and $s=7/6$. 
The compression modulus is $K=200\; {\rm MeV}$.
The evolution given by (\ref{e1}) is deterministic,
fluctuations appear only due to numerical noise\cite{ColBur93}.
We are using  75
pseudo-particles per nucleon and a pseudo-particle width,
 $\sigma_r = 0.53 \;{\rm fm}$, is adjusted so that the 
isovector giant dipole energy is reproduced at a single mass number. The experimental behavior of centroid energy with mass number is than reproduced. We have checked different numbers of test particles. The dependence of observables on the width is discussed in \cite{MW98}.
Numerically the ground state of nucleons is realized by Wood- Saxon shapes of density and Fermi spheres in momentum.
%

The distribution $n_a({\bf p})=\int d{\bf r} a_i f({\bf p,r})$ of mass distribution, $a_i=1$, isospin, $a_i=\tau_i$, kinetic energy, $a_i=\frac{p_i^2}{2m}$, and kinetic isospin energy, $a_i=\frac{\tau_ip_i^2}{2m}$, is
\begin{eqnarray}
n_a(p,\vartheta,\varphi) &=& \sum_{i=1}^{A N} \frac{(2\pi)^3}{N}  a_i
  \frac{\delta(p-p_i)}{p_i^2} \delta(\varphi-\varphi_i)
  \frac{\delta(\vartheta-\vartheta_i)}{\sin(\vartheta_i)} \;.
\end{eqnarray}
Radial integration determines a spherical distribution
\begin{eqnarray}
\bar{n}_a(\vartheta,\varphi) &=& \sum_{i=1}^{A N} \frac{1}{N}  a_i
  \delta(\varphi-\varphi_i) \frac{\delta(\vartheta-\vartheta_i)}
{\sin(\vartheta_i)},
\end{eqnarray} 
which can be decomposed into spherical harmonics
\begin{eqnarray}
\bar{n}_a(\vartheta,\varphi) &=& \sum_{l=0}^{\infty} \sum_{m=-l}^{l} 
  a_{lm} Y_{lm}(\vartheta,\varphi)\;, \\
 a_{lm} &=& \sum_{i=1}^{A N} \frac{1}{N}  a_i 
 Y_{lm}^\star(\vartheta_i,\varphi_i).
\end{eqnarray}
The observable distributions $\bar{n}_a(\vartheta,\varphi)$ are normalized 
to $\sqrt{4\pi} \; a_{00}$, i.e. to mass number $A$, total isospin $T$,
kinetic energy $E_{kin}$ and kinetic isospin energy $E_{kinT}$, respectively. 

As a measure for the strength of the resonances we use
\begin{eqnarray}
\hat{n}_a(\vartheta) = \int_0^{2\pi} d\varphi \, 
\bar{n}_a(\vartheta,\varphi) &=&
\sum_{l=0}^{\infty} a_{l0} {\textstyle \sqrt{\frac{2l+1}{4\pi}}} 
P_l(\cos\vartheta)
\end{eqnarray} 
with Legendre polynomials $P_l$.
The amplitudes of multipole moment, $a_{l0} \sqrt{\frac{2l+1}{4\pi}}$, 
will be displayed
as a function of time in the following graphs.
The value $a_{10}$ means the dipole moment, vanishing for isoscalar resonances,
$a_{20}$ is characterizing the quadrupole oscillations and $a_{30}$
the octupole ones.

\begin{figure}[h]
\epsfxsize=\hsize \epsfbox{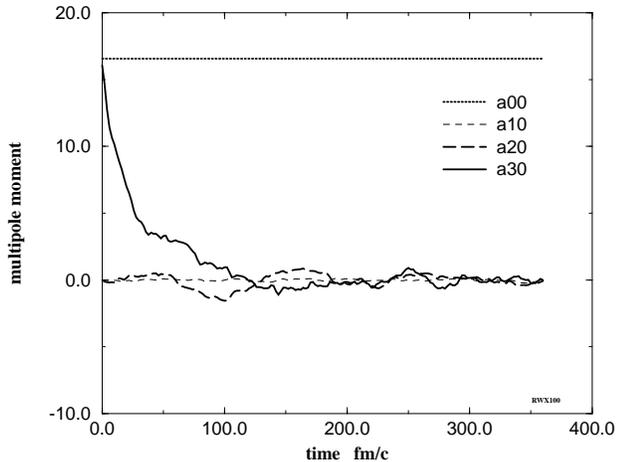}
\caption{The picture shows the time evolution of multipole moments
  $\protect\sqrt{\frac{2l+1}{4\pi}} a_{l0}(t), l=0,1,2,3$ of mass
  density concerning to normalization, dipole, quadrupole and octupole
  oscillation. Excitation is due to
 modification in momentum space only.\label{X100}} 
\end{figure}

\begin{figure}[h]
\epsfxsize=\hsize \epsfbox{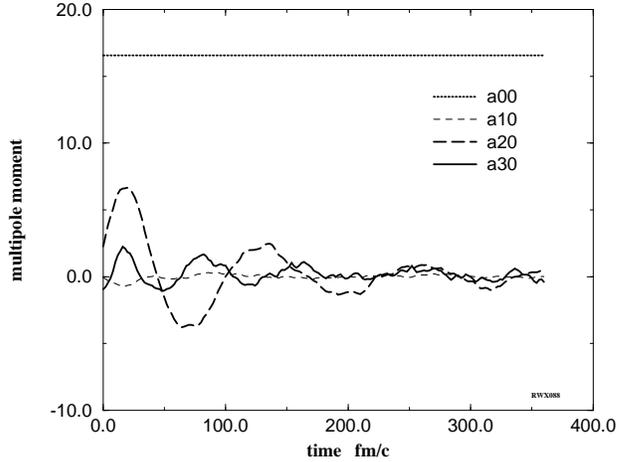}
\caption{This picture shows the time evolution of multipole moments
  $\protect\sqrt{\frac{2l+1}{4\pi}} a_{l0}(t), l=0,1,2,3$ of mass
  density. Excitation is due to
 asymmetric splitting (0.3 to 0.7) of the nucleus in two pieces using
 a plane in coordinate space. Then both pieces are accelerated towards each
 other.\label{X088}}
\end{figure}

%
As a first initial condition we use the ground state distribution of coordinates while the momentum distribution is deformed anisotropically.
We have modified the momenta in a way which corresponds to a giant octupole mode. The local densities and currents remain the same as in ground state.
Figure \ref{X100} shows that at start time $t=0$ there is a pure
giant octupole, which is
damped out and a quadrupole resonance developes instead. The monopole and dipole amplitudes which should remain constant document the stability of simulation.
In agreement with the already mentioned hypothesis, the
octupole mode is over-damped. The figure shows the nonlinear behavior of mode coupling. Within the linear response the damping rate is expected to be independent of initial conditions. We choose now other initial conditions to show that the result is very much dependent on initial conditions. 
Therefore we
split the nucleus in two parts of mass ratio 3:7 in accordance with symmetry of the octupole oscillation 
and accelerate both pieces towards each other. Experimentally it
might be realized as a central collision of two nuclei with
corresponding masses.
In figure \ref{X088} a clear quadrupole resonance appears and also 
a smaller octupole resonance can be seen. Both are damped out.
Consequently there is no evidence for an over-damped octupole mode in this case.

In order to
understand the different initializations, we split
the kinetic energy into a thermal part and a collective part
according to 
$\langle {\bf p}^2\rangle=\left\langle({\bf p}-\langle {\bf p}\rangle)^2\right\rangle +
\langle {\bf p}\rangle^2$.
We analyze the time
development of the total collective energy in the system
\begin{eqnarray}
E_{coll}(t) &=& \frac{h^2}{2m} \int d{\bf r} \varrho({\bf r}) \langle {\bf p}\rangle^2({\bf r},t)
\label{ecol}
\end{eqnarray}
with the mean current
\begin{eqnarray}
\langle {\bf p}\rangle({\bf r},t) &=& \frac{1}{\varrho({\bf r})} \int \frac{d{\bf p}}{(2\pi)^3} \;{\bf p}
\; f({\bf p,r},t)
\end{eqnarray}
and the density
\begin{eqnarray}
\varrho({\bf r},t) &=& \int \frac{d{\bf p}}{(2\pi)^3} \; f({\bf p,r},t) \; .
\end{eqnarray}

\begin{figure}[h]
\epsfxsize=\hsize \epsfbox{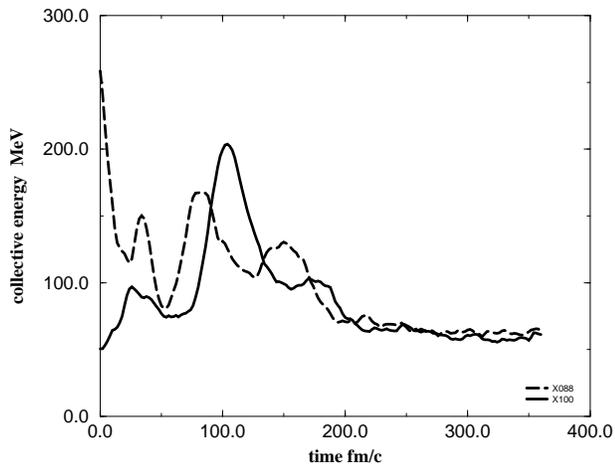}
\caption{The time evolution of collective energy. The solid line
 corresponds to the excitation scheme of figure \ref{X100} and the
 long dashed line corresponds to the figure
  \ref{X088}, respectively. While the first starts without collective
  energy, the second one starts with maximal collective energy.
\label{coll}}
\end{figure}

In figure \ref{coll} the development of collective energy can be
seen. There is a background of about 50~MeV due to fixed correlations
caused by finite width of pseudo-particles 
 as one can see from the following estimation. Using (\ref{2}) and (\ref{3}) in (\ref{ecol}) one obtains
\begin{eqnarray}
E_{coll}
   &\approx& {\textstyle \frac{1}{\varrho_0}} \sum_{i=1}^{AN} \sum_{j=1}^{AN}
   {\textstyle \frac{1}{N^2}} p_i p_j \nonumber\\
&&\times\int dr f_S(r-r_i(t),\sigma_r)
   f_s(r-r_j(t),\sigma_r)
\nonumber\\
&\approx& {\textstyle \frac{1}{\varrho_0}} \sum_{i=1}^{AN} \sum_{j=1}^{AN}
   {\textstyle \frac{1}{N^2}} p_i p_j f_S(r_i(t)-r_j(t),\sqrt{2}\sigma_r)  \nonumber\\
&\ge& {\textstyle \frac{1}{\varrho_0 N}
\frac{1}{(\sqrt{2\pi}\sqrt{2}\sigma_r)^3}} \sum_{i=1}^{AN} {\textstyle
\frac{1}{N}}
p_i.
\end{eqnarray}
For simulation parameter of $^{208}Pb$, 
$\rho_0=0.162\; {\rm fm}^{-3}$, $N=75$, 
$\sigma_r=0.53\;{\rm fm}$,
we obtain a basic collective energy of ${54 \rm MeV}$. Using more testparticles would diminish this level.

The solid line corresponding to
figure \ref{X100} shows no initial
collective energy. 
The exclusive initial excitation in momentum space without correlation
in spatial domain leads to zero initial collective energy.
This correlations are forming during time evolution.
Of course, there is no  center of mass motion, otherwise we would see
just the mean streaming velocity.
 
This situation is changed if we use the second preparation with simple
momentum--space--correlations. 
The long dashed line in figure
\ref{coll} shows initial collective correlations corresponding to
figure \ref{X088}.
Since we can deposit enough collective energy in this case, we
observe a clear octupole motion.

In order to compare with \cite{BloShi93} we calculate the adiabaticity
index $\eta$
defined in \cite{BloShi93} as the ratio of maximum radial
surface velocity to the maximum particle speed. A smaller ratio denotes a
more adiabatic shape changes in relation to the particle speed.
In analogy we define such index as a ratio
\begin{eqnarray}
\eta &=& \frac{\frac{\partial}{\partial t}\sqrt{<r^2>(\vartheta)}}{v_F}
\end{eqnarray}
of the root mean square radius speed in forward direction (opening angle
$\vartheta$) and the Fermi velocity.
With opening angle 0.4 rad we obtain a maximum $\eta=0.12$ for figure \ref{X100} and
$\eta=0.30$ for figure \ref{X088}.
This shows that we are essentially still in the adiabatic regime described in
\cite{BloShi93}. 

The nonlinear behavior described so far already documents that we are in a regime of large amplitude oscillations where linear response fails. The corresponding radius elongation
in coordinate space varies about 10 \%.

To summarize, we have observed  octupole resonances 
in finite nuclei dependent on
the initial configuration.
The appearance of an octupole mode was shown to be possible by 
correlating the spatial and momentum
initial excitation.
It is possible to
excite an octupole mode with sufficient collective energy deposited initially.
We suggest that isoscalar giant octupole resonances should be
possible to observe in nuclear collisions of mass ratio about 3:7 corresponding to octupole symmetry.
For a mass ratio of e.g.  
1:1 we observe no octupole resonance in simulation.

The fruitful discussions with Peter Schuck and P. Lipavsk\'y are gratefully acknowledged.
This work was supported by a grant (R.W.)
and project No. 44185 from the Max--Planck--Gesellschaft.



\end{document}